\documentclass{aa}  
\usepackage{natbib}
\bibpunct{(}{)}{;}{a}{}{,} 
\usepackage{graphicx}
\usepackage{txfonts}
\begin{document}
   \title{Gamma-ray emission from the solar halo and disk: a study with EGRET data}

   \subtitle{}

   \author{Elena Orlando
          \inst{1}
          \and
          Andrew W. Strong\inst{1}
          }

   \offprints{elena.orlando@mpe.mpg.de}

   \institute{Max-Planck-Institut f\"ur extraterrestrische Physik,
Postfach 1312, D-85741 Garching, Germany }

   \date{Received 10 October 2007; Accepted 11 January 2008 }

 
  \abstract
  {The Sun has recently been predicted to be an extended source of gamma-ray emission, produced by inverse-Compton (IC) scattering of cosmic-ray (CR) electrons on the solar radiation field. The emission was predicted to be extended and a confusing foreground for the diffuse extragalactic background even at large angular distances from the Sun. The solar disk is also expected to be a steady gamma-ray source. While these emissions are expected to be readily detectable in the future by GLAST, the situation for available EGRET data is more challenging. }
   {  The theory of gamma-ray emission from IC scattering on the solar radiation field by Galactic CR electrons is given in detail. This is used as the basis for detection and model verification using EGRET data.   }
{  We present a detailed study of the solar emission using the EGRET database, accounting for the effect of the emission from 3C 279, the moon, and other sources, which interfere with the solar emission. The analysis was performed for 2 energy ranges, above 300 MeV and for 100-300 MeV, as well as for the combination to improve the detection statistics. The technique was tested on the moon signal, with our results consistent with previous work. 
   }
{Analyzing the EGRET database, we find evidence of emission from the solar disk and its halo.   The observations are compared with our model for the extended emission. The spectrum of the solar disk emission and the spectrum of the extended emission have been obtained. The spectrum of the moon is also given.}
  { The observed intensity distribution and the flux are consistent with the predicted model of IC gamma-rays from the halo around the Sun.}


\titlerunning{Solar gamma-ray emission}
\maketitle
%

\section{Introduction}
Solar quiet gamma-ray astronomy started playing a significant role in the early 90's thanks to the EGRET mission. \citet{hudson} pointed out the importance of gamma-ray emission from the quiet Sun as an interesting possibility for highly sensitive instruments such as EGRET.  
\citet{seckel} estimated that the flux of gamma rays produced by cosmic-ray interactions on the solar surface would be detectable by EGRET. This emission is the result of particle cascades initiated in the solar atmosphere by Galactic cosmic rays. Meanwhile \citet{thompson97} detected in the EGRET data the gamma rays produced by cosmic ray interactions with the lunar surface. Although they expected similar interactions on the Sun, the EGRET data they analyzed did not show any excess of gamma rays consistent with the position of the Sun. They obtained an upper limit of the flux above 100 MeV of 2 $\times$ 10$^{-7}$cm$^{-2}$s$^{-1}$ . \citet{fairbairn}, analyzing the EGRET data of the solar occultation of 3C279 in 1991, found a photon flux of about 6 $\times$ 10$^{-7}$ cm$^{-2}$s$^{-1}$ above 100 MeV in the direction of the occulted source, which they used to put limits on axion models; however they did not consider the extended solar emission.

Inverse Compton scattering of cosmic-ray electrons on the solar photon halo around the Sun has been estimated to be an important source of gamma-ray emission  \citep{orlando,moskalenko} \footnote{The same process from nearby stars has been studied in \citet{orlando} and \citet{orlandoc}}. In our previous work, we predicted this to be comparable to the diffuse background even at large angular distance from the Sun. The formalism has now been improved and more accurate estimates have been obtained. The anisotropic scattering formulation and the solar modulation have been implemented, in order to give a more precise model for the EGRET data. In this work we explain in detail our model of the extended emission produced via inverse Compton scattering of cosmic-ray electrons on the solar radiation field.  

A first report of our detection with EGRET data was given in \citet{orlandob}. Here we describe the full analysis including the spectrum of disk and extended components.  Our result is very promising for the forthcoming GLAST mission which certainly will be able to detect the emission even at larger angular separation from the Sun. Moreover, the extended emission from the Sun has to be taken into account since it can be strong enough to be a confusing background for Galactic and extragalactic diffuse emission studies.


\section{Theoretical model}
Inverse-Compton scattering of solar optical photons by $\sim$ GeV cosmic-ray electrons produces $\gamma$-radiation with energies of 100 MeV and above.
Improving on the approximate estimates given in \citet{orlando}, in this paper the anisotropic formulation of the Klein-Nishina cross section has been introduced, taking into account the radial distribution of the photons propagating outward from the Sun. Our elegant analytical formulation has been replaced by a numerical solution depending on  the angle between the momenta of the cosmic ray electrons and the incoming photons. As pointed out in  \citet{orlando} this will affect the results at about the 10$\%$ level.  Moreover, while in our previous work we used the modulated cosmic-ray electron spectrum as observed at earth, here a formulation of the solar modulation as a function of the distance from the Sun has been taking into account.

The gamma-ray intensity spectrum is:
\begin{equation}
\label{eq1}
I(E_{\gamma})=\frac{1}{4\pi}\int\epsilon(E_{\gamma})dx
\end{equation}
where the emissivity $\epsilon$ is given by:
\begin{eqnarray}
\epsilon(E_{\gamma})=\int dE_{e} \times
\nonumber\\
\times \int \sigma_{K-N}(\gamma,E_{ph},E_{\gamma})~ n_{ph}(E_{ph},r)~c~N(E_{e},r)~dE_{ph}
\label{eq2}
\end{eqnarray}
$N(E_{e},r)$ is the electron density, with $E_{e}$ electron energy, $n_{ph}(E_{ph},r)$ the solar photon density as function of the distance from the Sun $r$ and the solar photon field $E_{ph}$, $\sigma$ is the Klein-Nishina cross section and $\gamma=E_{e}/m_{e}$.
The cosmic-ray electrons have been assumed isotropically distributed everywhere in the heliosphere.
In \citet{orlando}, an elegant analytical formulation in eq (\ref{eq3}) was obtained for the isotropic case, assuming a simple inverse square law for the photon density for all distances from the Sun and a cosmic ray electron spectrum considered constant for all distances and determined from experimental data (see Fig.(\ref{fig4})) for 1 AU.
In order to obtain the inverse Compton radiation over a line of sight at an angle $\alpha$ from the Sun, the photon field variation over the line of sight has to be known. 
\begin{figure}[!t]
	\includegraphics[width=9cm, angle=0]{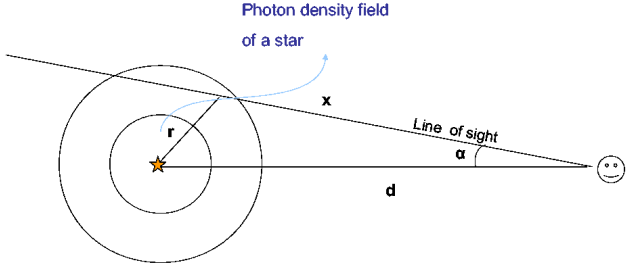}
	\caption{Definition of variables for eq.(\ref{eq3}) describing the geometry of the solar inverse Compton emission.}
	\label{fig1}	
\end{figure}
As shown in Fig.(\ref{fig1}), for a given point $x$ on the line of sight, the surface photon density is proportional to $1/r^{2}$, where $r^{2}=x^{2}+d^{2}-2$~$x$~$d$ $\cdot$ $cos~\alpha$, with $d$ distance from the Sun. Integrating the photon density over the line of sight from $x=0$ to $x=\infty$ one obtains:
{\setlength\arraycolsep{2pt}
\begin{eqnarray}
n_{ph}(E_{ph})~R^{2}\int_{0}^{\infty}\frac{dx}{x^{2}+d^{2}-2~x~d \cdot cos~\alpha}=
\nonumber\\
=-n_{ph}(E_{ph})~R^{2}~arctan\left( \frac{d \cdot cos~\alpha-x}{d \cdot sin~\alpha}\right) /(d \cdot sin~\alpha~)=
\nonumber\\
=n_{ph}(E_{ph})~R^{2}\left( \frac{ \pi/2+arctan(cot~\alpha)}{d \cdot sin~\alpha} \right) 
\label{eq3}
\end{eqnarray}}
which depends only on the angle $\alpha$, where $R$ is the solar radius and $n_{ph}(E_{ph})$ is the solar photon density. For large distance from the source $n_{ph}(E_{ph})$=1/4 $n_{BB}(E_{ph})$, where $n_{BB}(E_{ph})$ is the black-body density of the Sun.
Integrating over solid angle and using eq.(\ref{eq1}) and eq.(\ref{eq2}) with the Klein-Nishina cross section, the total photon flux produced by inverse Compton scattering within an angle $\alpha$ becomes:  
{\setlength\arraycolsep{2pt}
\begin{eqnarray}
I(E_{\gamma})=\frac{1}{4\pi} \int_{0}^{2\pi} d\varphi \int_{0}^{\alpha} sin~\alpha~ d \alpha \int dE_{ph} {}
\nonumber\\
{}\times \int \sigma_{K-N}(\gamma,E_{ph}, E_{\gamma})~c~N(E_{e})~dE_{e}~ n_{ph}(E_{ph})~R^{2} \times {}
\nonumber\\
{}\int_{0}^{\infty}\frac{dx}{x^{2}+d^{2}-2~x~d \cdot cos~\alpha}= {}
\nonumber\\
{}=\frac{R^{2}}{16~d}\left( \pi \alpha + (\frac{\pi}{2})^{2}-arctan^{2}~(cot~\alpha)\right) {}
\nonumber\\
{} \times \int dE_{ph} \int \sigma_{KN}~c~N(E_{e})~n_{BB}(E_{ph})~dE_{e} {}
\label{eq4}
\end{eqnarray}
which for small $\alpha$ is proportional to $\alpha/d$ and the intensity $I$ (per solid angle) is proportional to 1/($\alpha d$).
For the case of the anisotropic formulation of the Klein-Nishina cross section and for the electron modulation along the line of sight, we have to use numerical computations, already adopted in \citet{orlandob}.

\subsection{Solar photon field}
\subsubsection{Basic relations}
The Sun is treated as black body where the energy density is characterized by the effective temperature on the surface (T=5777 K) following the Stephan-Boltzmann equation. 
For the photon density close to the Sun, the simple inverse square law is inappropriate. In this work the emission from an extended source has been evaluated. 
The distribution of photon density from the Sun, as extended source, is given by integrating over the solid angle with the variables shown in Fig.(\ref{fig2})  
\begin{figure}[!h]
 \centering
	\includegraphics[width=.5\textwidth, angle=0]{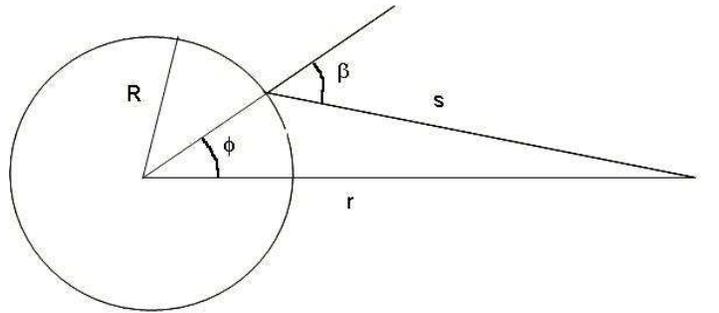}	
	\caption{ Variables involved in eq.(\ref{eq7}) for the calculation of the photon density around the Sun, where R is the solar radius and r is shown also in Fig. (\ref{fig1})}
	 \label{fig2}
\end{figure}
{\setlength\arraycolsep{2pt}
\begin{eqnarray}
n_{ph}(E_{ph},r)=n_{BB}(E_{ph})\int_{0}^{\phi_{MAX}} \frac{2\pi~R^{2}~sin~\phi ~cos~\beta ~d\phi}{4\pi~s^{2}}
\nonumber\\
=0.5 ~n_{BB}(E_{ph}) \left[ 1-\sqrt{1-(R/r)^{2}}\right]
\label{eq5}
\end{eqnarray}
with $cos~\beta$=$(r~cos~\phi~-~R)/s$ and $cos~\phi_{MAX}~=~R$/$r$.
For large distances from the Sun it reduces to the inverse square law.

\subsubsection{Deviations of solar spectrum from black body}
We checked the effect of deviations of the solar spectrum from a black body.
We used the black-body approximation with the AM0 solar spectrum \footnote{taken from http://www.spacewx.com/}, referring to 2007, period of solar minimum, extrapolated back to the Sun. Since the estimated inverse Compton emission from the Sun is not affected significantly ($<$1$\%$) considering the actual spectrum, we decided to keep the black body approximation for simplicity.
We also verify that the EUV and XUV range of the solar spectrum do not affect the inverse-Compton emission. The irradiance at the shortest wavelengths varies significantly with solar conditions and the range of variability spans a few percent at the longest wavelengths, from 40\% to more than 500\% across the EUV and XUV , from solar minimum to solar maximum. 
Moreover, during solar flare, XUV increases, reaching factors of eight or more, and EUV increases on the order of 10-40\% \citep{eparvier}. However, \citet{oncica} found that the X-ray flares cannot contribute to the total solar irradiance fluctuations and even the most energetic X-ray flares cannot account for more than 1 mW/$m^{2}$.
We found that the estimated inverse-Compton emission from the Sun is not affected significantly by the XUV/EUV flux and its variability.

\subsection{Comparison of isotropic/anisotropic formulations} 
In \citet{orlando} the isotropic formulation for the Klein-Nishina cross section given in \citet{moskalenkoa}  
 was used in the analytical formulation for estimating the IC gamma-ray emission .
In the present analysis we used the following anisotropic Klein-Nishina cross section in a more convenient form than in \citet{moskalenkoa}:

{\setlength\arraycolsep{2pt}
\begin{eqnarray}
\sigma_{K-N}(\gamma,E_{ph}, E_{\gamma}) =  
\frac{ \pi r_{e}^{2}m_{e}^{2}}{E_{ph} E_{e}^{2}}\times
\nonumber\\
  \left[ \left(  \frac{m_{e}}{E'_{ph}}\right) ^{2} (\frac{v}{1-v})^{2} {} -2 \frac{m_{e}}{E'_{ph}} \frac{v}{(1-v)}+(1-v)+\frac{1}{1-v} \right]  
\label{eq6}
\end{eqnarray}
where 
$v=E_{\gamma}/E_{e}$ and
$\gamma$=$E_{e}/m_{e}$, with $m_{e}$ electron mass
and 
\begin{equation}
E'_{ph}=\gamma E_{ph}(1+cos~ \eta)
\label{eq7}
\end{equation}
where $\eta$ is the scattering angle for the relativistic case, with $\eta=0$ for a head-on collision.
In Fig.(\ref{fig3}) is shown the contribution to the flux as a function of distance along the line-of-sight for the isotropic and anisotropic cross-sections. 
\begin{figure}[!h]
\centering
\includegraphics[width=.35\textwidth, angle=270]{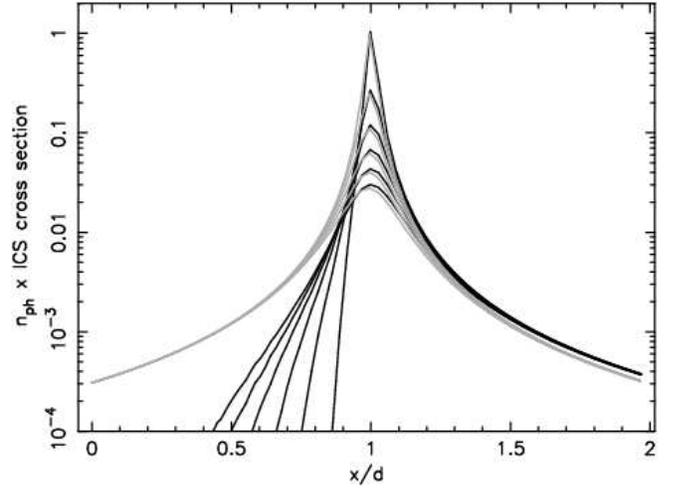}
\caption{ Relative contribution (arbitrary units) of the isotropic (gray lines)
and anisotropic (black lines) cross sections, multiplied by the photon
density along the line of sight for E$_{e}$=10$^{4}$ MeV and E$_{\gamma}$=100 MeV. $x$ and $d$
are defined in Fig 1. The lines represent different angles from the Sun,
from 1$^{\circ}$ to 6$^{\circ}$ (right to left).}.
\label{fig3}
\end{figure}

The integrated inverse-Compton emission was compared for isotropic and anisotropic cross sections; the maximum difference of the intensity for the two Klein-Nishina cross sections is of about 15$\%$ around 300 MeV and for 0.5$^{\circ}$ angular distances from the Sun.

\subsection{Electron spectrum}
The solar modulation of cosmic ray electrons has been widely studied, but is still not completely understood.  
\citet{mueller} analyzed the data of Helios-1 from 1974 to 1975, a period that corresponds to the approach to solar minimum,  and reported the measurements of proton and helium cosmic ray intensities for the 20-50 MeV/n energy interval between 0.3 and 1 AU. The proton gradients they obtained were small and generally consistent with zero, while the helium gradients were positive but small. On the basis of Helios-1 data, \citep{kunow} stated: `The radial variation seems to be very small for all particles and all energy ranges'.  The study of \citep{seckel} was based on this hypothesis. Since that time, the modulation theories have been developed including effects like drifts and current sheet tilt variations \citep{thomas} and short time variations. In this work we care only about long period modulations for solar maximum and minimum and we assume that the longitude gradient is negligible.
\citet{mcdonald}, analyzing cycle 20 and 22 of solar minimum between 15 and 72 AU, found that most of the modulation in this period occurs near the termination shock (assumed to be at 100 AU). On the other hand, from solar minimum to solar maximum the modulation increases mainly inside the termination shock. Recently \citet{morales}, investigating the radial intensity gradients of cosmic rays from 1 AU to the distant heliosphere and interpreting the data from IMP8, Voyagers 1 and 2, Pioneer 10 and BESS for Cycles 21, 22 and 23, found different radial gradients in the inner heliosphere compared to \citet{mcdonald}. In this region they obtained an average radial gradient of $\simeq$ 3\% /AU for 175 MeV H and $\simeq$ 2.2\% /AU for 265MeV/n He, that, at 1AU, give an intensity smaller than \citet{mcdonald} for H and higher for He. \citet{gieseler} analyzed the data from Ulysses from 1997 to 2006 at 5AU. They found a radial gradient of 4.5\%/AU for $\alpha$ particles with energies from 125 to 200 MeV/n, which is consistent with previous measurements.

In this work, our model of the modulation of cosmic rays uses the formulation given in \citet{gleeson}, using the studies 
regarding the radial distribution of cosmic rays in the heliosphere at solar minimum and maximum.
The cosmic ray electron spectrum is given by the well known force field approximation of cosmic ray nuclei used to obtain the modulated differential intensity J(r,E) at energy E and distance r from the Sun \citep{gleeson}. As demonstrated by  \citet{caballero} it is a good approximation for Galactic cosmic rays in the inner heliosphere. The differential intensity of the modulated spectrum is given by:
\begin{equation}
J(r, E)=~J ( \infty, ~E+ ~\Phi(r) ) ~ \frac{E(E+2E_{0})}{(E+~\Phi(r)+2E_{0})(E+\Phi(r))} 
\label{eq8}
\end{equation}
where E is kinetic energy in MeV and $E_{0}$ is the electron rest mass.
We use the local interstellar electron spectrum $J ( \infty, ~E+ ~\Phi(r))$.
$\Phi(r)=(Ze/A)\phi$ where $\phi$ is the modulation potential, Z the charge number and A the mass number.
In order to compute the modulation potential in the inner heliosphere, we used the parameterization found by \citet{fujii} for Cycle 21 from 100 AU to 1 AU, neglecting the time dependence and normalizing at 1 AU for 500 MV (solar minimum) and 1000 MV (solar maximum). As \citet{moskalenko} we find:
\begin{equation}
\Phi(r)=\Phi_{0}(r^{-0.1}-r_{b}^{-0.1})/(1-r_{b}^{0.1})
\label{eq9}
\end{equation}
where $\Phi_{0}$ is the modulation potential at 1 AU, of 500 and 1000 MV for solar minimum and maximum respectively;  $r_{b} =100 AU$ and $r$ is the distance from the Sun in AU. 
Figure (\ref{fig4}) shows the local interstellar electron spectrum \citep{strong} and the modulated spectrum at 1 AU, compared with the data.
\begin{figure}[!h]
 \centering
	\includegraphics[width=.35\textwidth, angle=270]{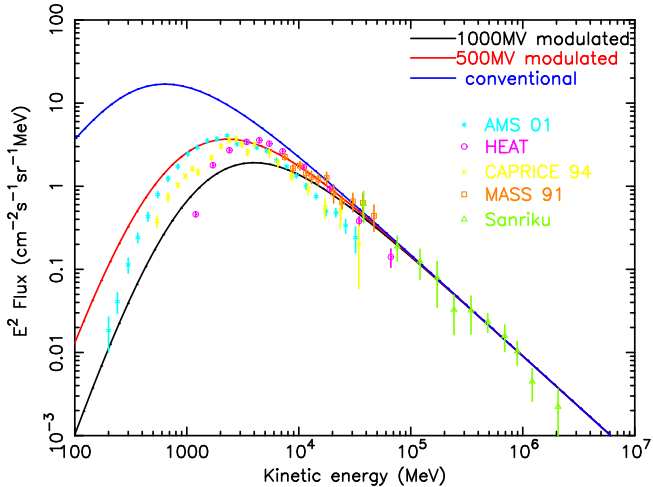}
	
	\caption{ Local interstellar electron spectrum (blue line) as in \citet{strong} and the modulated at 1 AU of 500 MV (red line) and 1000 MV (black line) compared with the data. See \citet{strong} for data reference. }
	 \label{fig4}
\end{figure}
The electron intensity as a function of distance from the Sun is given in Fig.(\ref{fig5}) for different energies and for modulation potential 500 MV. 
\begin{figure}[!h]
 \centering
	\includegraphics[width=.45\textwidth, angle=0]{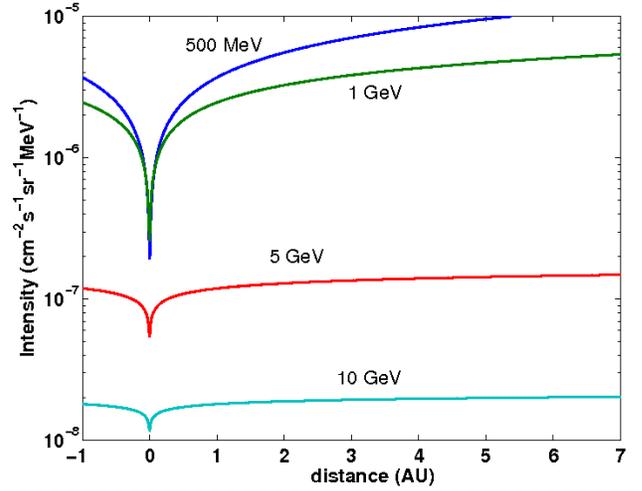}
	
	\caption{ Electron intensity as a function of distance from the Sun in AU, where 0 corresponds to the Sun, for different energies and for modulation potential $\Phi_{0}=500 MV$. }
	\label{fig5}
\end{figure}

\begin{figure}[!h]
 \centering
	\includegraphics[width=.36\textwidth, angle=270]{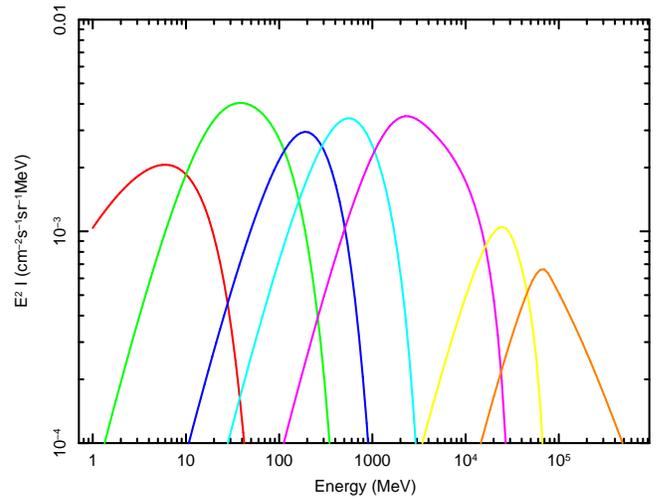}
	
	\caption{ Contribution to the IC emission for different electron energy ranges, left to right: $100-10^{3}$ MeV, $(1-3) \times 10^{3}$ MeV, $(3-5) \times 10^{3}$ MeV, $(0.5-1) \times 10^{4}$ MeV,$(1-5) \times 10^{4}$ MeV, $(0.5-1) \times 10^{5}$ MeV, $1 \times 10^{5}- 5 \times 10^{7}$ MeV.  }
	 \label{fig6}
\end{figure}
The major contribution to the energy range 100-500 MeV we are interested in comes from electrons with energy between 1 and 10 GeV, as shown in Fig. (\ref{fig6}), obtained for the local interstellar electron spectrum, with no modulation.
To compute the inverse-Compton extended emission from the heliosphere and compare it with the EGRET data, details of the solar atmosphere, the non-isotropic solar wind and the asymmetries of the magnetic field have been neglected, considering the limited sensitivity of EGRET.   
Since the biggest uncertaintes in the gamma-ray emission come from the cosmic-ray electron spectrum close to the Sun, in our model we considered two possible configurations of the solar modulation within 1 AU. The "naive" approximation is to assume that the cosmic-ray flux towards the Sun equals the observed flux at Earth, since there is evidence that modulation by the solar wind does not significantly alter the spectrum once cosmic rays have penetrated as far as Earth. Moreover, high energy electrons are less sensitive to the modulation. This approximation gives an upper limit of the modelled flux.  The other approach is to assume that the electron spectrum varies due to solar wind effects within 1 AU. With this "nominal" approximation we assume that the formulation for solar modulation from 100 AU to 1AU can be extrapolated also below 1 AU, using eq. (\ref{eq8}). This gives an approximate lower limit in our model. 

\subsection{Calculated extended solar emission}
Figure (\ref{fig7}) shows the spectrum of the emission for two different angular distances from the Sun, without modulation and for two levels of solar modulation ($\Phi$=500, 1000 MV, respectively for solar minimum and solar maximum) and for the cases of upper and lower limits described above. 

\begin{figure}[!h]
 \centering
	\includegraphics[width=.45\textwidth, angle=0]{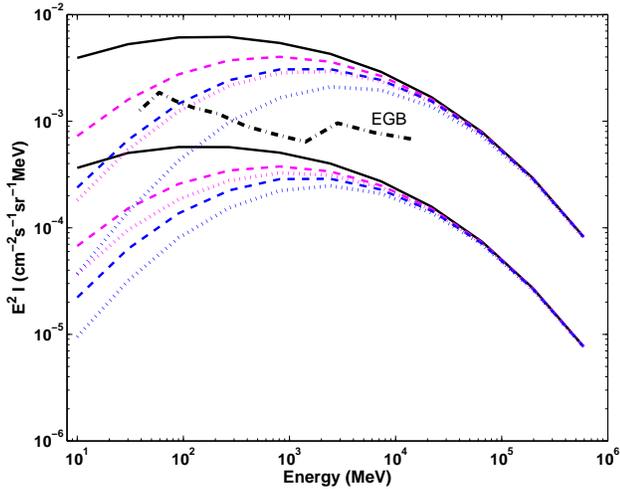}
	\caption{ Spectrum of the emission for (top to bottom) 0.5$^{\circ}$ and 5$^{\circ}$ angular distances from the Sun and for different conditions of solar modulation. Solid lines: no modulation, pink lines: $\Phi_{0}$=500MV, blue lines:$\Phi_{0}$=1000MV, dashed lines: naive model, dotted lines: nominal model. EGB is the extragalactic background as in \citet{stronga}}
	 \label{fig7}
\end{figure}
The angular profile of the emission is shown in Fig.(\ref{fig8}) above 100 MeV without modulation, for two levels of solar modulation ($\Phi$=500, 1000 MV) and for the cases naive and nominal. The emission is extended and is important compared to the extragalactic background (around 10$^{-5}$cm$^{-2}$s$^{-1}$sr$^{-1}$) even at very large angles from the Sun. Even around 10$^{\circ}$ it is still about 10$\%$ of the extragalactic background and with the senstivity of GLAST should be included in the whole sky diffuse emission.
\begin{figure}[!h]
 \centering
	\includegraphics[width=.45\textwidth, angle=0]{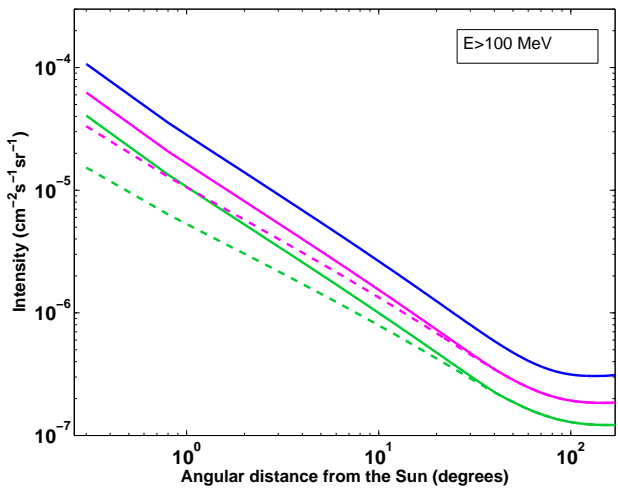}
	\caption{ Angular profile of the emission as a function of the angular distance from the Sun above 100 MeV. Blue line: no modulation, pink lines: $\Phi_{0}$=500MV, green lines: $\Phi_{0}$=1000MV, solid lines: naive model, dashed lines: nominal model.}
	 \label{fig8}
\end{figure}
An example of the intensity of the inverse-Compton emission predicted in a region of 10$^{\circ}$ from the Sun is shown in Fig.(\ref{fig9}) for 300-500 MeV.
\begin{figure}[h!]
    \centering
\includegraphics[width=.3\textwidth, angle=0]{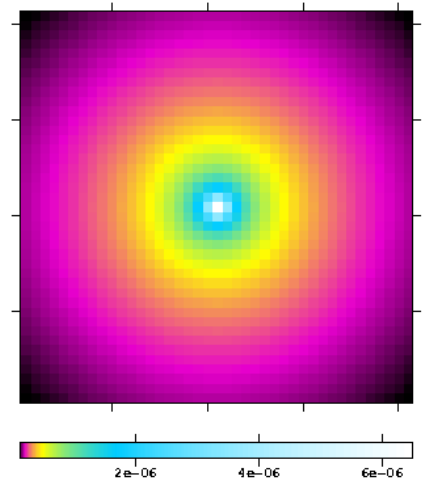}
 
    \caption{Inverse-Compton emission modelled for a region of 10$^{\circ}$ from the Sun for 300-500 MeV and for the naive model of 1000 MV modulation. Intensity is given in cm$^{-2}$s$^{-1}$sr$^{-1}$. In the figure, the minimum value of the intensity is 1.2 $\times$ 10$^{-7}$ cm$^{-2}$s$^{-1}$sr$^{-1}$.}
\label{fig9}
\end{figure}
While the inverse Compton emission is expected to be readily detectable in future by GLAST, the situation for available EGRET data is more challenging. In the following section we present our study with the EGRET database.

\section{EGRET data preparation and selection}
We analyzed the EGRET data using the code developed for the moving target such as Earth \citep{petry}. The software permits us to produce images centred on a moving source and traces of other sources around with respect to the centred source.
To perform the analysis for the Sun, we analyzed the data in a Sun-centred system. 
In order to have a better sensitivity and hence detection of our emission component, the diffuse background was reduced by excluding the Galactic plane within 15$^{\circ}$ latitude. Otherwise all available exposure from October 1991 to June 1995 was used (viewing periods 1, 11, 12, 19, 26, 209, 221, 320, 325, 410, 420), excluding the big solar flare in June 1991. When the Sun passed by other gamma-ray sources (moon, 3C 279 and several quasars), these sources were included in the analysis. In order to obtain the flux of the moon to be fitted, the same method performed for the Sun has been made with all data moon centred. Details will be given in \citet{petrya}.

\section{EGRET analysis}
We fitted the EGRET data in the Sun-centred system using a multi-parameter likelihood fitting technique (see Appendix).  The region used for fitting is a circle of radius 10$^{\circ}$ around
the Sun. 
The total number of counts predicted includes inverse Compton and solar disk flux, the moon, QSO 3C279, other background sources and the background. We left as free
parameters the solar extended inverse-Compton flux from the model, the solar disk flux (treated as point source since the solar disk is not resolvable by EGRET), a
uniform background, and the flux of 3C279 - the dominant background point source. The moon flux was determined
from moon-centred fits and the 3EG source fluxes were fixed at their catalogue values. All components were
convolved with the energy-dependent EGRET PSF.
\subsection{Model of extended solar emission}
The solar extended inverse-Compton emission was modelled within 10$^{\circ}$ radius from the Sun and with a pixel size of 0.5$^{\circ}$. The model was convolved with the energy-dependent EGRET PSF and implemented in the fit, with a scale factor as free parameter. We took into account the different approximations to the modulation described above, since the data we used cover a period from October 1991 when there was a solar maximum to June 1995 close to the solar minimum. Since the emission we are looking for is very close to the EGRET sensitivity limit, we had to take as much exposure as possible and it was not possible to perform an analysis splitting the data according to solar modulation. 
\subsection{3C279}
Since 3C 279 is the brightest source that passes close to the Sun, we decided to leave its flux as free parameter in the fit. We fixed the spectral index at 1.96, the value given in the EGRET catalogue. The flux obtained was then compared to that in the EGRET catalogue.
\subsection{{\bf Other background sources}}
Many sources passing within 10$^{\circ}$ from the Sun were included in the analysis. We decided to implement their fluxes as fixed parameters. With the code \citep{petry} it was possible to know the exact viewing period of any source and hence to determine its flux from the 3rd EGRET catalogue \citep{hartmann}. Hence, we took the flux value in the EGRET catalogue corresponding to the period we used, when this was listed. Since the catalogue does not contain the flux values corresponding to all the periods in which the sources were close to the Sun, in case of incertainty we decided to use the first entry in the catalogue, which is the one from which the source position was derived. In almost all cases, this is the detection with the highest statistical significance, which in most cases corresponds to the maximum value of the flux. We also took the spectral index from the EGRET catalogue. The values adopted for the sources are listed in Table (\ref{table1}).
Since the spectral index of J2321-0328 is not known, the fit was performed without this source above 300 MeV, since in the literature there is no evidence of its detection above this energy. In order to verify that this choice does not affect significantly the fit result, we tested also a typical spectral index of 2. Both fitting results  will be reported in the following results section. The traces of the background quasars were added together in the same data file, rescaled according to their fluxes. 

\begin{table}
\begin{minipage}[t]{\columnwidth}
\caption{Parameters of the background sources used for the analysis.}
\label{table1}
\centering
\renewcommand{\footnoterule}{}  
\begin{tabular}{l l l }
\hline \hline
Source& Flux ($>$100 MeV)\footnote{Units 10$^{-7}$ cm$^{-2}$s$^{-1}$}     &  spectral index     \\

\hline
 J0204+1458   &2.36  &     2.23   \\
 J0215+123   &1.80&     2.03  \\
 J1230-0204   &1.13  &     2.85   \\
 J1235+0233   &1.24  &     2.39   \\
 J1246-0651   &1.29  &     2.73   \\
 J1310-0517   &1.04  &     2.34   \\
 J1409-0745  &2.74  &     2.29   \\
 J2321-0328    &3.82 &     - \footnote{The choice of the spectral index we used is explained in the text} \\
\hline
\end{tabular}
\end{minipage}
\end{table}

\subsection{Diffuse background}
Since the count maps used for our analysis include only data above $\mid$b$\mid$ $>$15$^{\circ}$ latitude, and the Galactic plane is excluded, the Galactic emission can be well approximated by an isotropic background. We left its intensity as a free parameter in the fitting analysis.
\subsection{Moon}
The flux of the moon was determined from moon-centred data, fitting the flux and an isotropic background. Since the moon moves quickly across the sky, all other sources moving by are taken into account as a constant component included in the isotropic background.  The fit was performed for 2 energy ranges: 100-300 MeV and $>$300 MeV. The value of flux of the moon for the different energy ranges was obtained by maximum likelihood.  
The maximum likelihood ratio statistic used for this analysis is described in the appendix.
The values of the best fit fluxes and 1$\sigma$ errors are given in Table (\ref{table2}).
\begin{table}
\caption{Fitting results for the moon}             
\label{table2}      
\centering                          
\begin{tabular}{l l l l }        
\hline\hline                 
Energy (MeV)& Moon flux ($cm^{-2}s^{-1}$) & Background ($cm^{-2}s^{-1}sr^{-1}$)  \\    
\hline                        
   $> 100$ & (5.55~$\pm$~0.65)~ $\times 10^{-7}$ & 3.47 $\times 10^{-5}$ \\      
   $> 300$ & (5.76~$\pm$~1.66)~ $\times 10^{-8}$ & 1.01 $\times 10^{-5}$ \\
   $100-300$ & (4.98~$\pm$~0.57)~ $\times 10^{-7}$ & 2.13 $\times 10^{-5}$ \\
\hline                                   
\end{tabular}
\end{table}
Previous studies of the EGRET data (Thompson et al.) gave a flux of ($5.4 \pm 0.7) \times 10^{-7}$ cm$^{-2}$s$^{-1}$ above 100 MeV in agreement with our analysis. 
\begin{figure}[!h]
 \centering
	\includegraphics[width=.45\textwidth, angle=0]{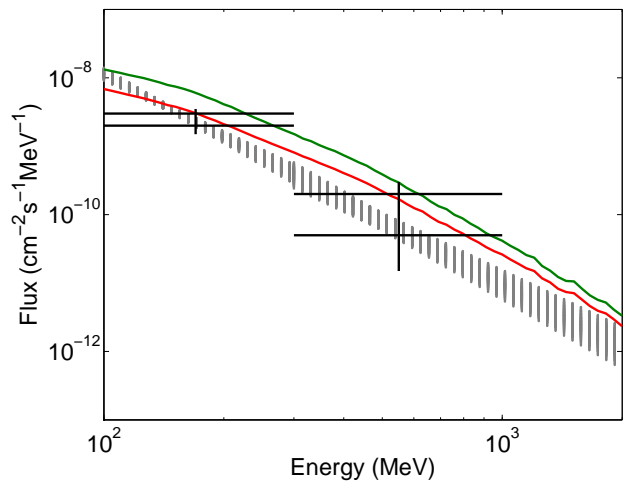}
	
	\caption{Spectrum of the moon obtained by the fitting analysis of the EGRET data between October '91 to June '95 (dashed region). Thickness of the line shows the error bars. Black bars are obtained by \citet{thompson97} for different solar conditions, while lines are the theoretical model of \citet{moskalenkob}, for solar maximum (red line) and solar minimum (green line).}
	 \label{fig10}
\end{figure}
The spectrum of the moon shown in Fig.(\ref{fig10}) has been obtained imposing a constant spectral index $\gamma$, 
extrapolated up to 2 GeV.

The lunar spectral index was determined as $3.05^{+0.38}_{-0.29}$. Errors are calculated from the uncertainties on the integrated fluxes.
Since the EGRET data we used to analyze the emission from the Sun extend from solar maximum to minimum, in this analysis we estimated an average lunar flux between the two periods of solar modulation. Comparing our spectrum with the model of \citet{moskalenkob} for different solar conditions, we find a good agreement at 100 MeV, while at higher energies the model is about a factor of two higher. More extensive analysis will be given in \citet{petrya}.

\section{Solar analysis results}

The analysis was performed for 2 energy ranges : 100-300 MeV and above 300 MeV as well as the combination to improve the detection statistics. This choice was determined by the limitations of the EGRET data.  Fig.\ref{fig11} shows  the smoothed count maps, centred on the Sun, for those energy ranges and for a region of 20$^{\circ}$ side. The fitting region has a radius of 10$^{\circ}$ centred on the Sun. Since the interesting parameters are solar disk source and extended emission, the likelihood is maximized over the other components.  The analysis was performed for four different estimates of the solar modulation.

\begin{figure}[!h]
 \centering
	\includegraphics[width=.25\textwidth, angle=0]{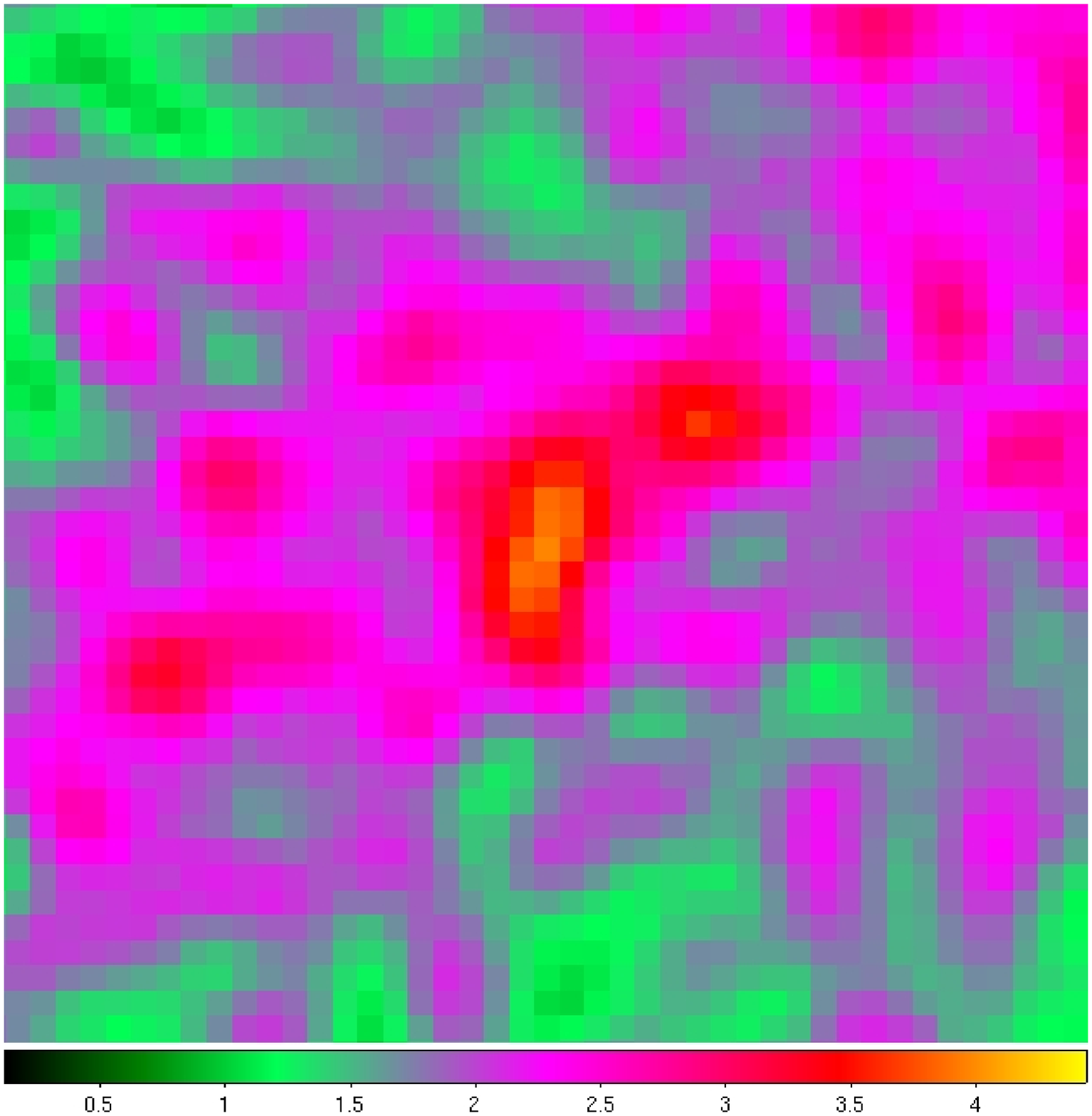}\\
	\includegraphics[width=.25\textwidth, angle=0]{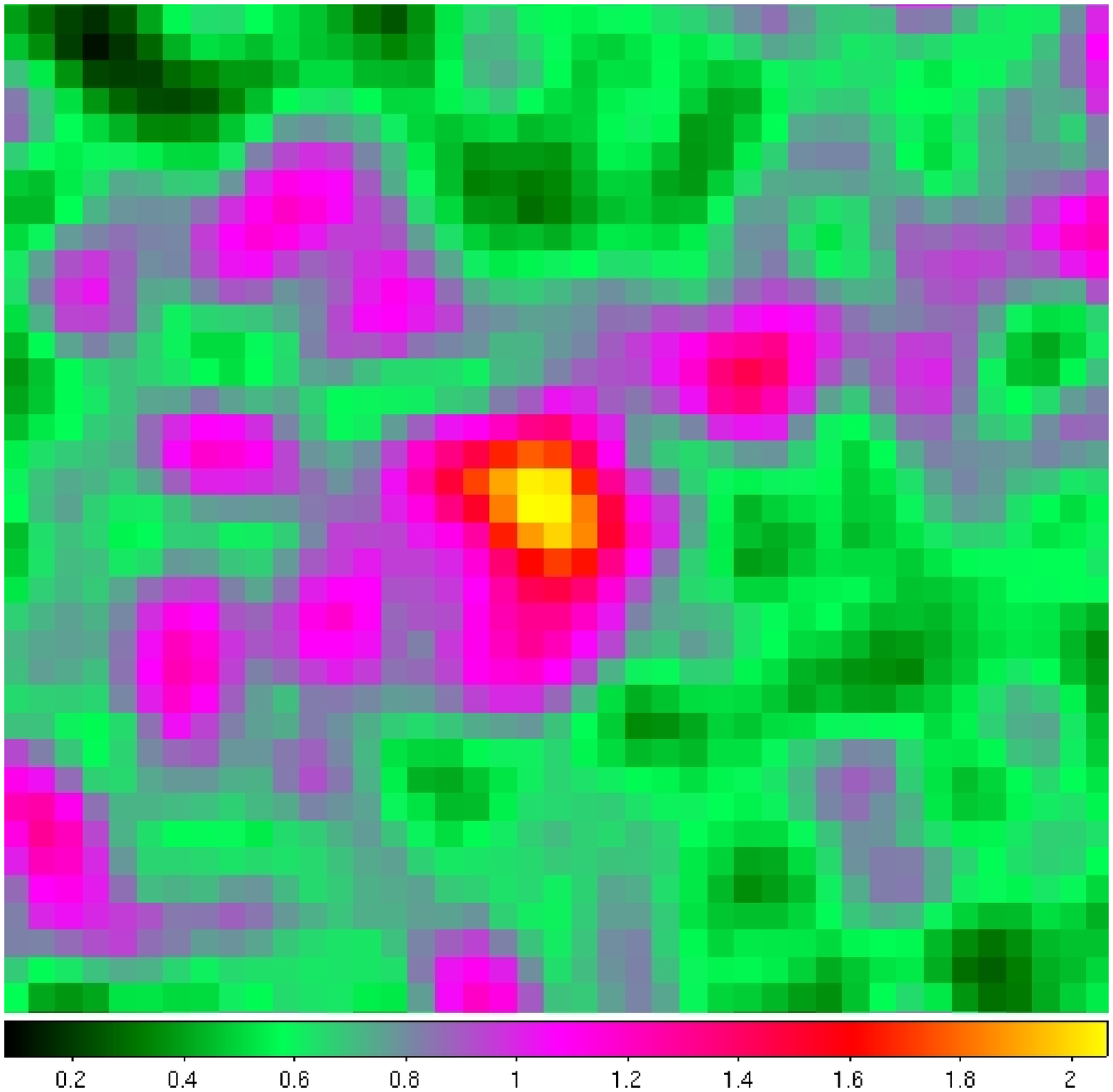}\\	
	\includegraphics[width=.25\textwidth, angle=0]{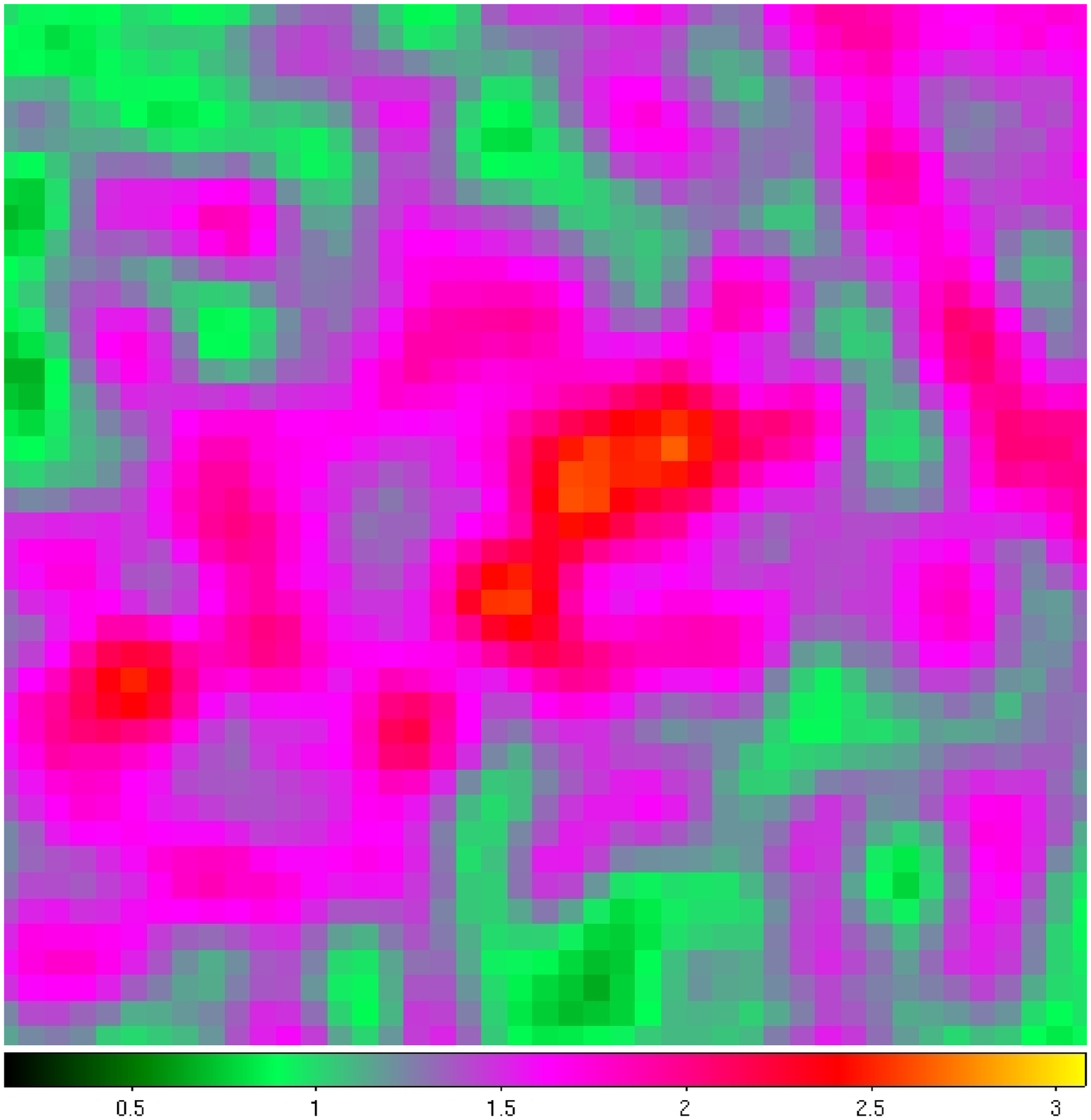}	%
\caption{EGRET Sun-centred counts maps (top to bottom)$>$100 MeV, $>$300 MeV and 100-300 MeV. The colorbar shows the counts per pixel. The area is 20$^{\circ}$ on a side and the maps are Gaussian smoothed to 3$^{\circ}$.}
	 \label{fig11}
\end{figure}

\begin{table*}
\caption{Results for different energy ranges and for the naive model of 1000 MV solar modulation. Fluxes are given in cm$^{-2}$s$^{-1}$ and intensities in cm$^{-2}$s$^{-1}$sr$^{-1}$. For details see text. }
\label{table3}      
\centering          
\begin{tabular}{l l l l l l }     
\hline\hline       
Model& &  $>$100 MeV & $>$300 MeV with J2321-0328 & $>$300 MeV no J2321-0328 &100-300 MeV\\
\hline\hline
   Source flux&best fit&  4.68$^{+4.17}_{-3.82}$ $\times 10^{-8}$ &  3.38$^{+1.80}_{-1.81}$ $\times 10^{-8}$& 3.38$^{+1.88}_{-1.84}$ $\times 10^{-8}$ &14.04  $\times 10^{-8}$  \\  
 &mean&  (5.42$\pm$3.20)$\times 10^{-8}$ &  (3.03$\pm$1.10)$\times 10^{-8}$& (3.65$\pm$1.71)$\times 10^{-8}$ &(14.2$\pm$9.1)$\times 10^{-8}$ \\ 
  &counts & 28 & 20 & 20 &39\\
\hline
   Extended flux&best fit & 3.83$^{+2.78}_{-2.80}$ $\times 10^{-7}$& 1.54$^{+1.06}_{-1.18}$ $\times 10^{-7}$&1.75$^{+1.08}_{-1.13}$ $\times 10^{-7}$&1.15 $\times 10^{-7}$\\
&mean & (3.89$\pm$2.16)$\times 10^{-7}$& (1.49$\pm$0.67)$\times 10^{-7}$&(1.70$\pm$0.87)$\times 10^{-7}$&(2.07$\pm$1.34)$\times 10^{-7}$\\
&   counts & 273&116&132& 83\\
\hline
Extended solar model&flux &2.18 $\times 10^{-7}$& 0.90 $\times 10^{-7}$&0.90 $\times 10^{-7}$&1.28 $\times 10^{-7}$\\
\hline
Background& intensity & 3.50$\times 10^{-5}$&1.13$\times 10^{-5}$&1.12$\times 10^{-5}$&2.48$\times 10^{-5}$\\
&counts & 2220&750&741&1603\\
\hline
3C279& flux & 4.88$\times 10^{-7}$&1.30$\times 10^{-7}$&1.30$\times 10^{-7}$&5.85$\times 10^{-7}$\\
&counts & 103&29&29&79\\
\hline
Prob. (flux=0)&& 6.4$\times 10^{-3}$&2.3$\times 10^{-4}$& 9.7$\times 10^{-5}$&0.24\\
\hline
Strongest bkg source &flux& 3.82$\times 10^{-7}$ & 1.28$\times 10^{-7}$    & 0.63$\times 10^{-7}$ & 2.54$\times 10^{-7}$\\
Total bkg sources & counts   &113 & 27     & 21 & 66\\
\hline
Moon& flux& 5.55$\times 10^{-7}$ & 0.57  $\times 10^{-7}$  & 0.57$\times 10^{-7}$ & 4.98$\times 10^{-7}$\\
 & counts& 37 & 4     & 4 & 29\\
\hline                  
\end{tabular}
\end{table*}
For the different models of solar modulation, the expected values of inverse Compton emission are in agreement with the data within 1$\sigma$, with rather big errors bars. This also means that, because of the limited sensitivity of EGRET, it is not possible to prove which model better describes the data. We found probabilities corresponding to 2.7, 3.6/4 and 1$\sigma$ above 100 MeV, above 300 MeV (including/excluding J2321-0328) and for 100-300 MeV respectively for the significance of the detection.
However, we chose the naive model of 1000 MV solar modulation in order to derive the spectra of the extended and disk emission. This is for two reasons: first this model produces the highest value of the likelihood; second it should be the most realistic for the period '91-'95, since the data cover the total period of solar maximum and the upcoming solar minimum.
{\it Hence, we report only the results for the naive case of 1000 MV solar modulation.}
\begin{figure*}[!h]
 \centering
	\includegraphics[width=.45\textwidth, angle=0]{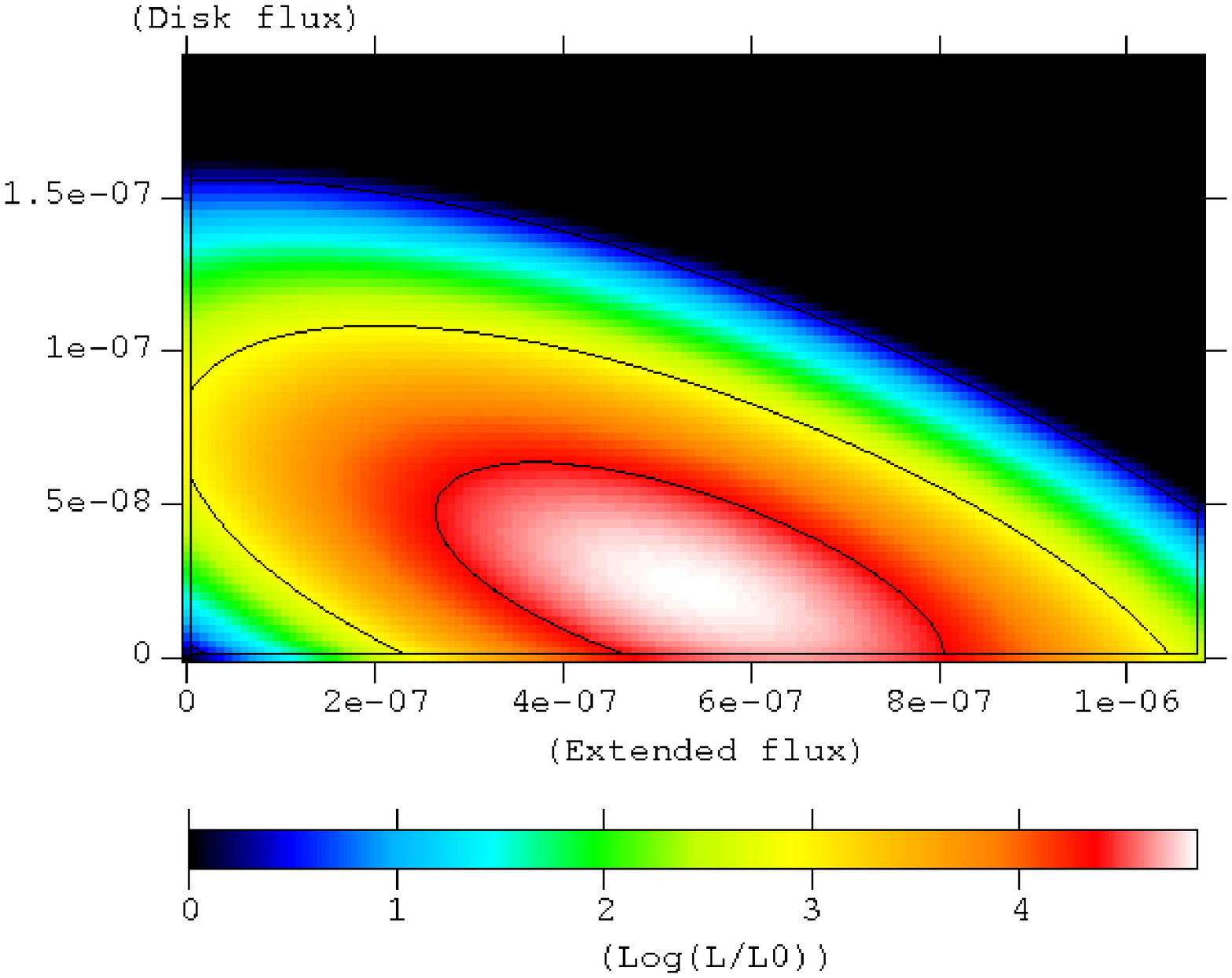}
	\includegraphics[width=.45\textwidth, angle=0]{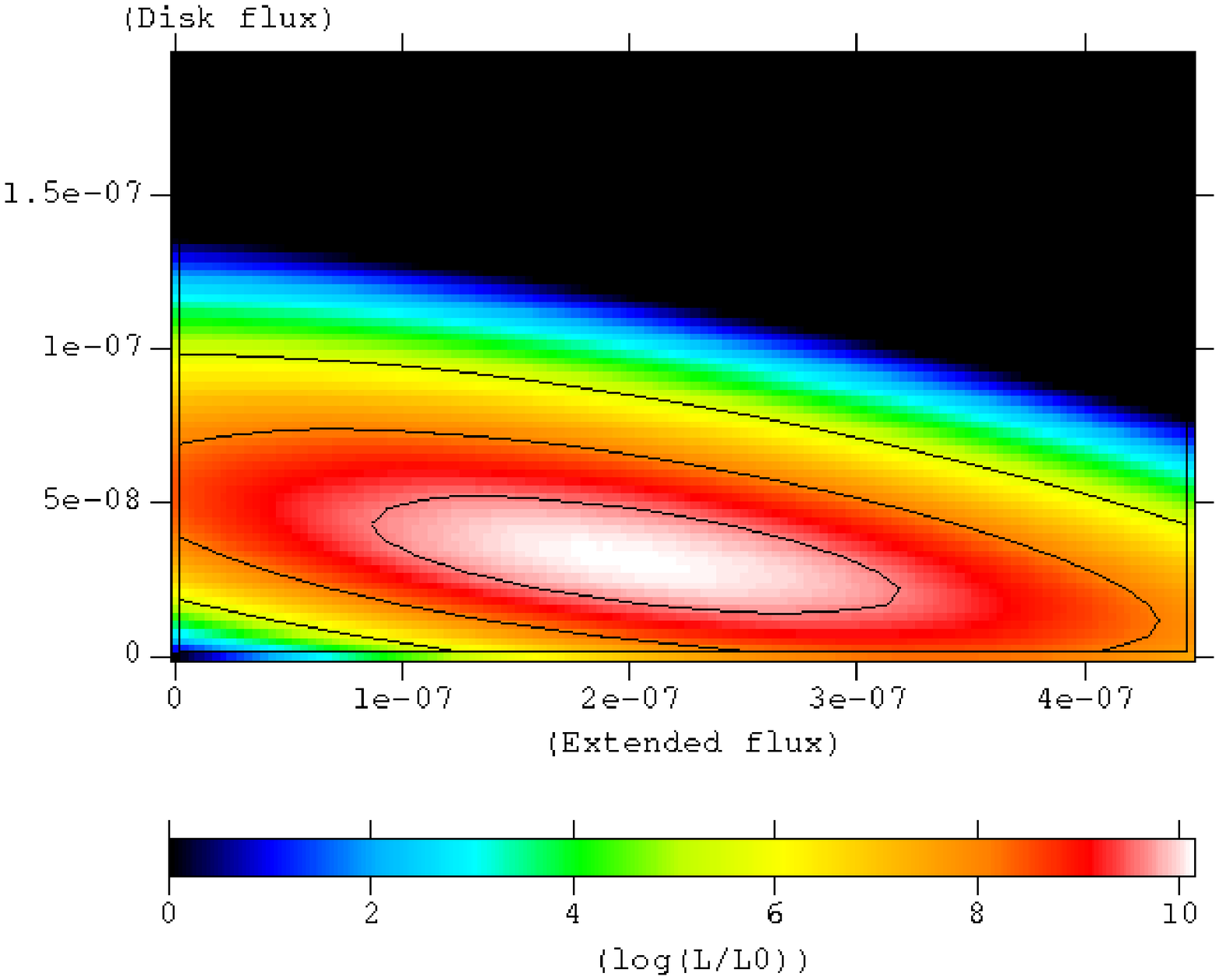}
	\includegraphics[width=.45\textwidth, angle=0]{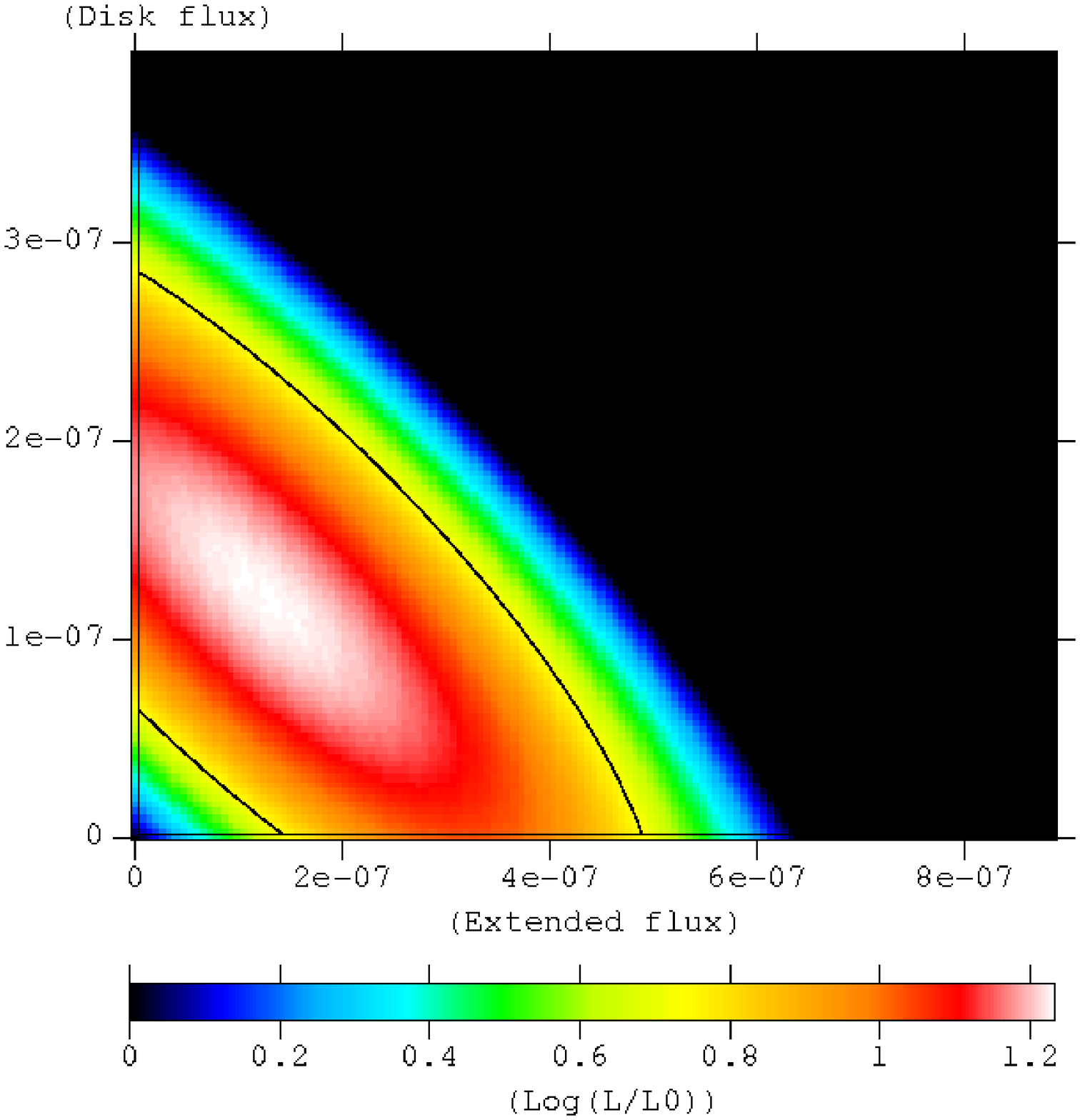}
	\caption{Logarithm of the likelihood ratio (ln~$(L/L_{0})$) as a function of the solar disk flux and an extended component for (top to bottom) $>$100 MeV, $>$300 MeV and 100-300 MeV . Color contours are different values of the ratio, as explained in the colorbar. Contours are obtained by allowing the background and the QSO 3C279 component to vary to maximize the likelihood for each value of disk flux and extended component. Contour lines define  1, 2 and 3$\sigma$  intervals for the 2 separate parameters.}
	 \label{fig12}
\end{figure*}
We used both frequentist and Bayesian approaches to analyze the results. The statistical methods are described in the appendix. 

Values of the best fit fluxes, 1$\sigma$ errors, counts and mean values with errors are given in Table \ref{table3}.
 Best fit values are calculated from the log-likelihood ratio statistic, while mean values are from the Bayesian method.
 Counts are for the maximum likelihood values, as are the background, the 3C279 fluxes and the probability of the null hypothesis. Values in the first column were used to have more statistics for detection, while separate energies are best for the analysis. For 100-300 MeV range error bars can not be determined by the frequentist method.
 We also give the probability of the null hypothesis (i.e. zero flux from disk and extended emission) using the likelihood ratio statistic. The table contains two cases above 300 MeV with different fluxes of the background sources. The first is obtained including the source J2321-0328 whose spectral index was fixed at a typical value of 2, since it is not given in the 3EG catalogue. The second case does not include that source, since it has not been detected at energy above 300 MeV. We took the second case for the following analysis. These two cases give also an estimate of the uncertainty of our results.
\begin{figure*}[!h]
 \centering
	\includegraphics[width=.32\textwidth, angle=0]{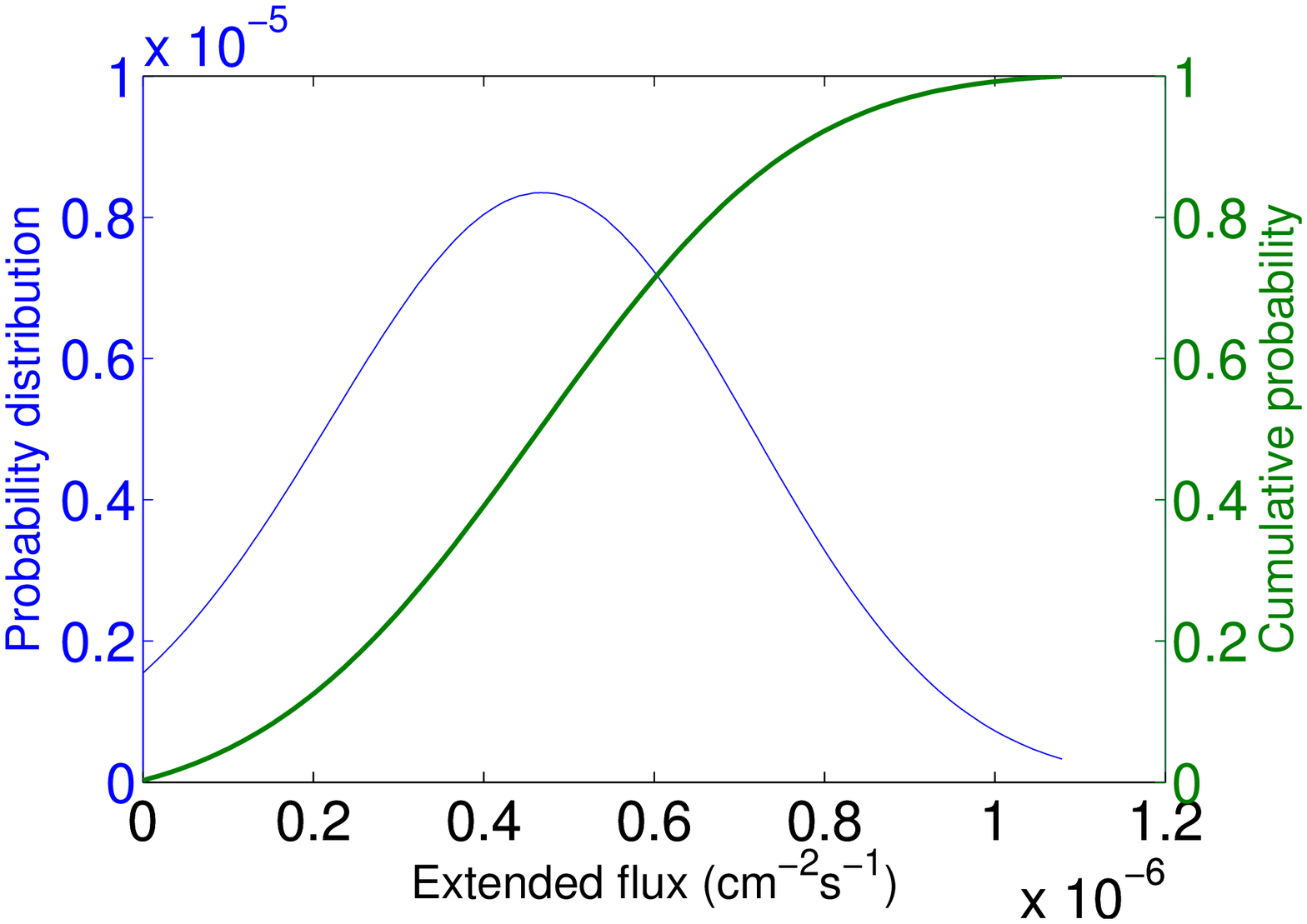}
	\includegraphics[width=.32\textwidth, angle=0]{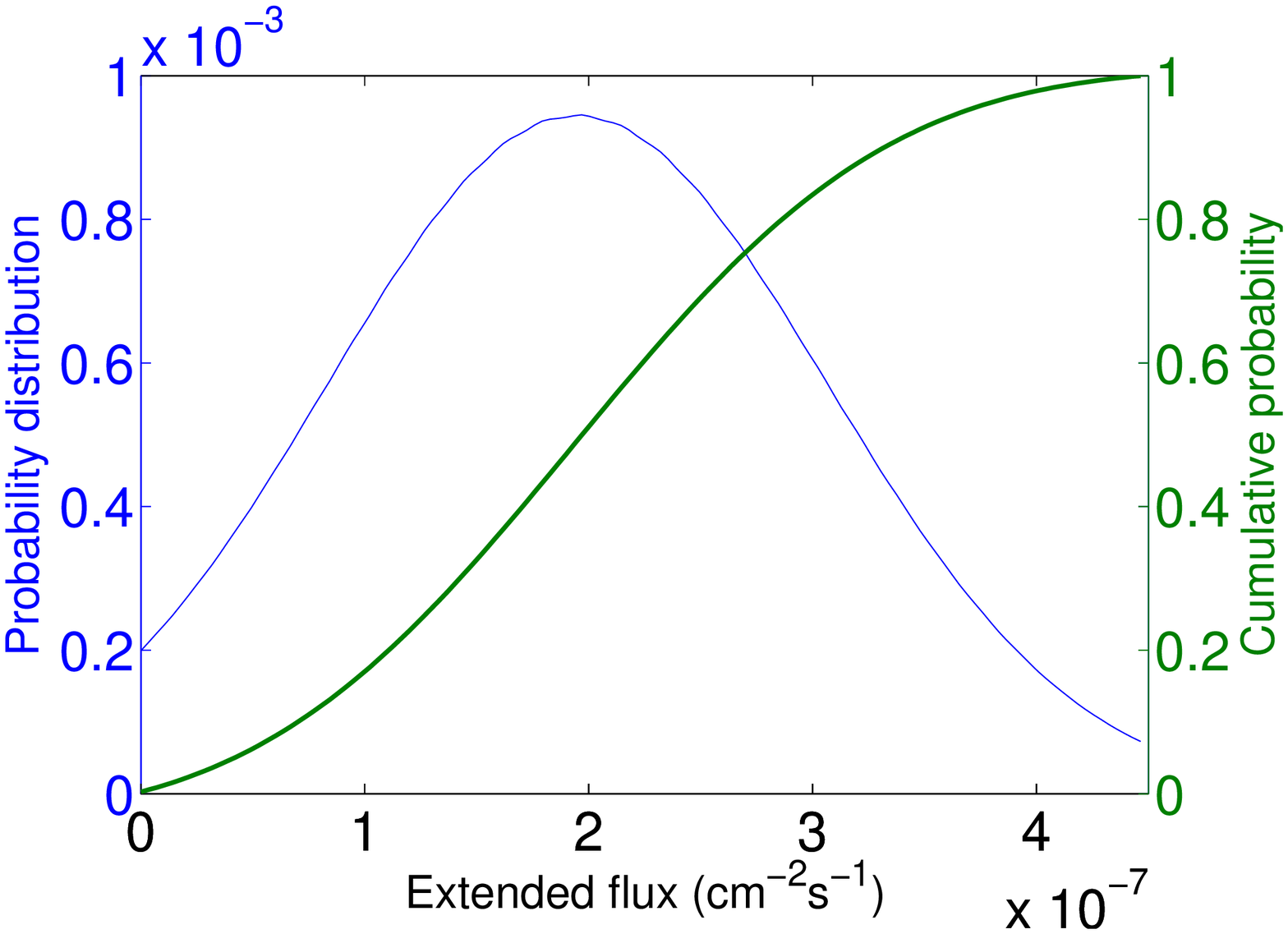}
	\includegraphics[width=.32\textwidth, angle=0]{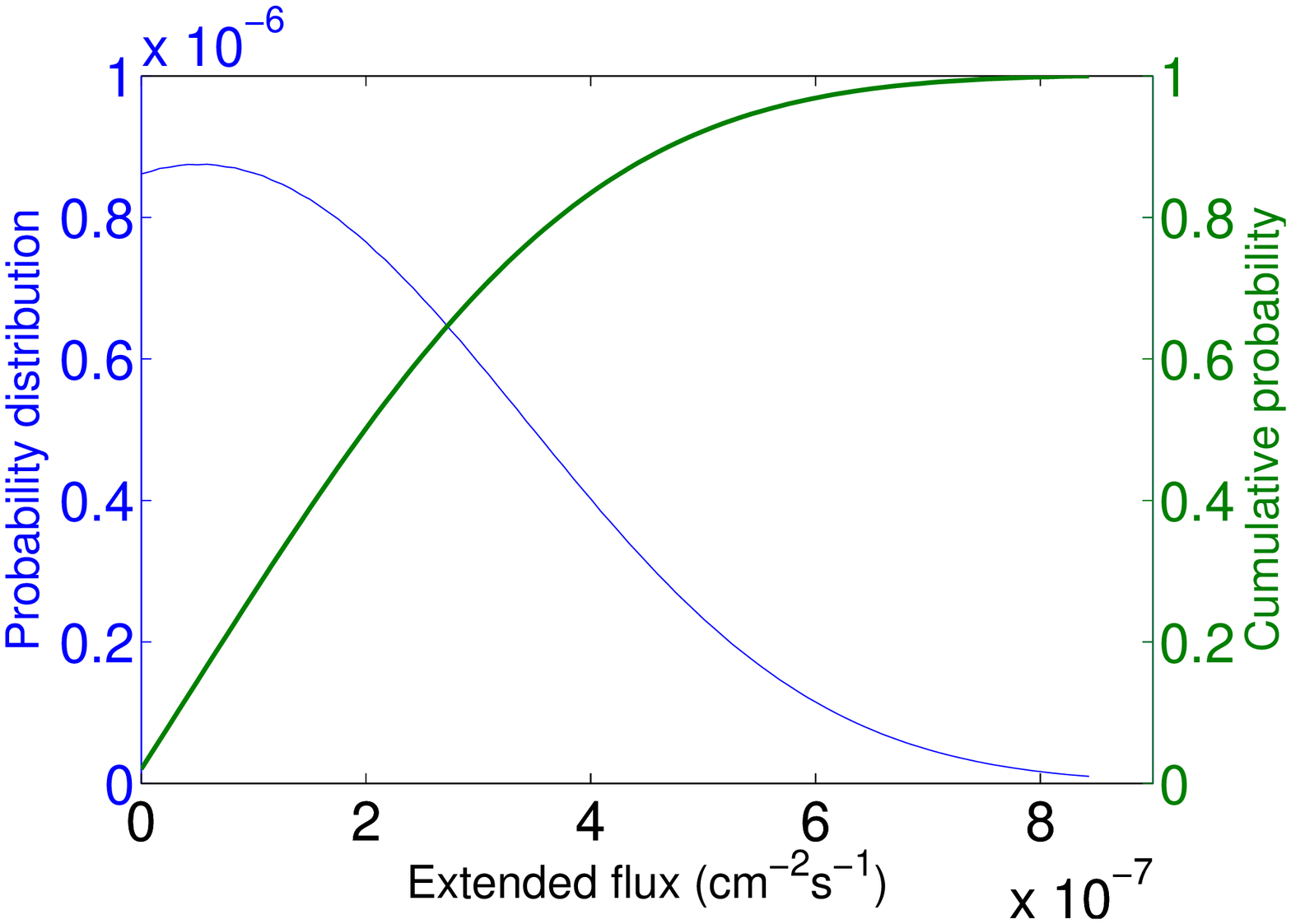}	
	\includegraphics[width=.32\textwidth, angle=0]{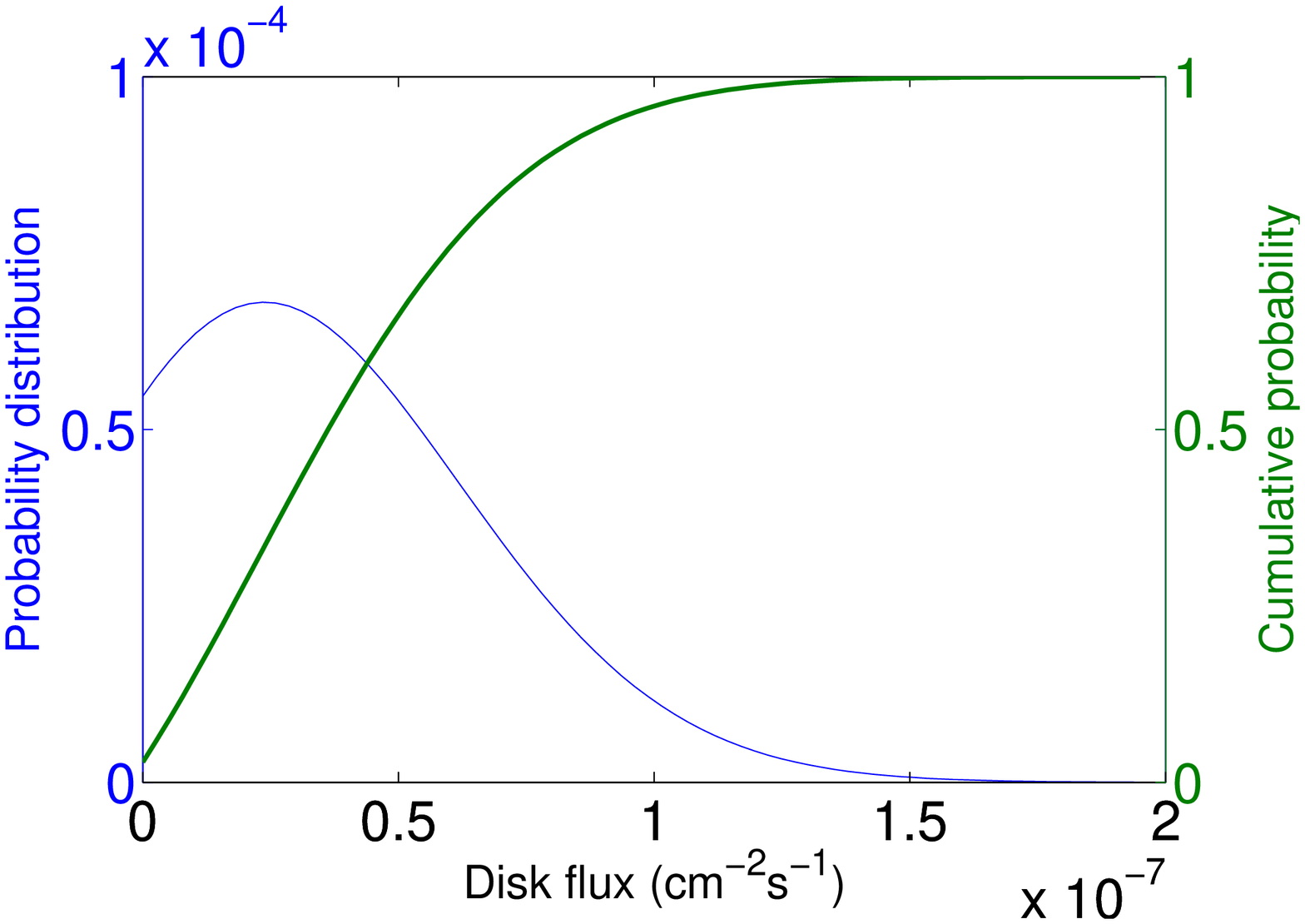}
	\includegraphics[width=.32\textwidth, angle=0]{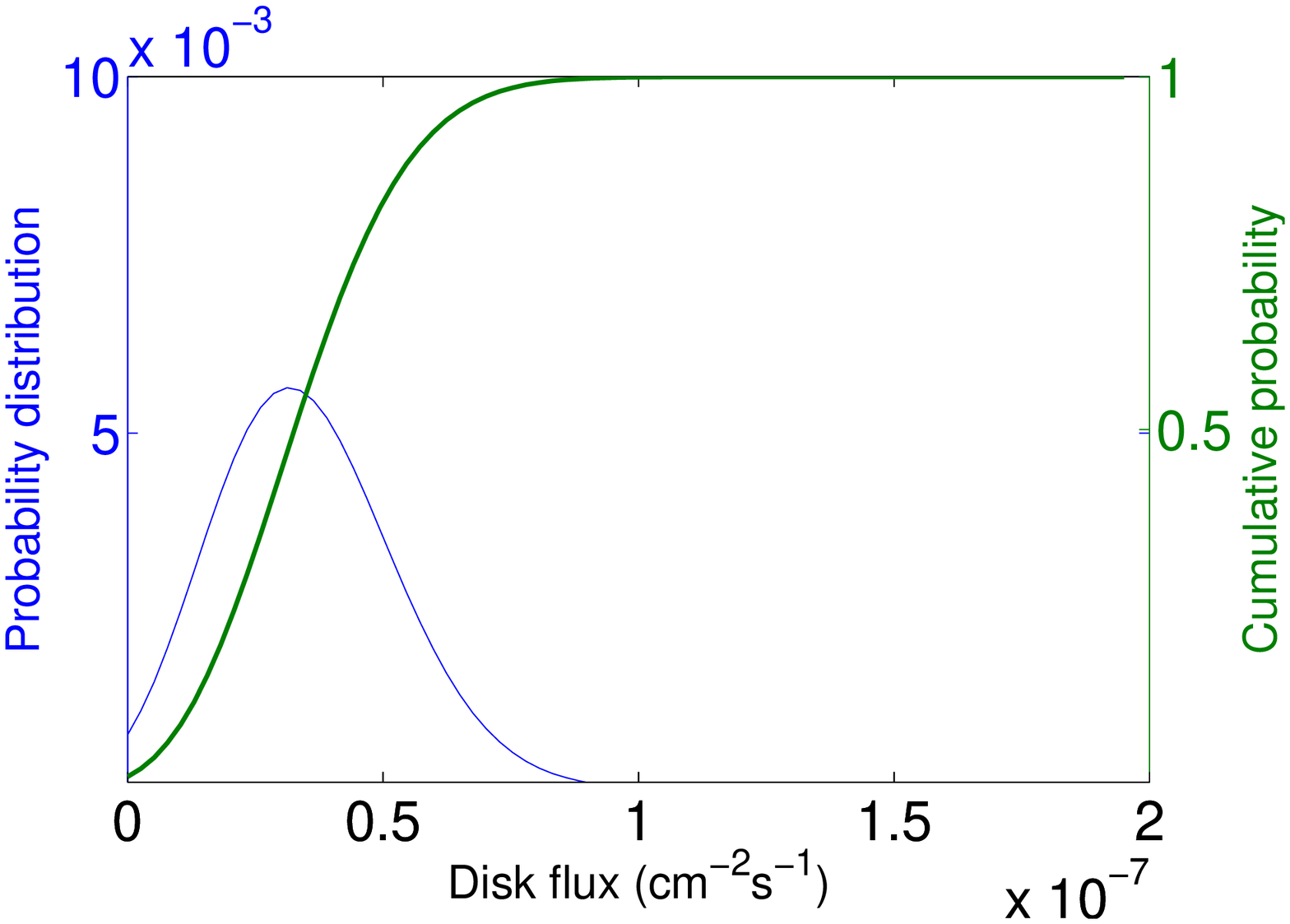}
	\includegraphics[width=.32\textwidth, angle=0]{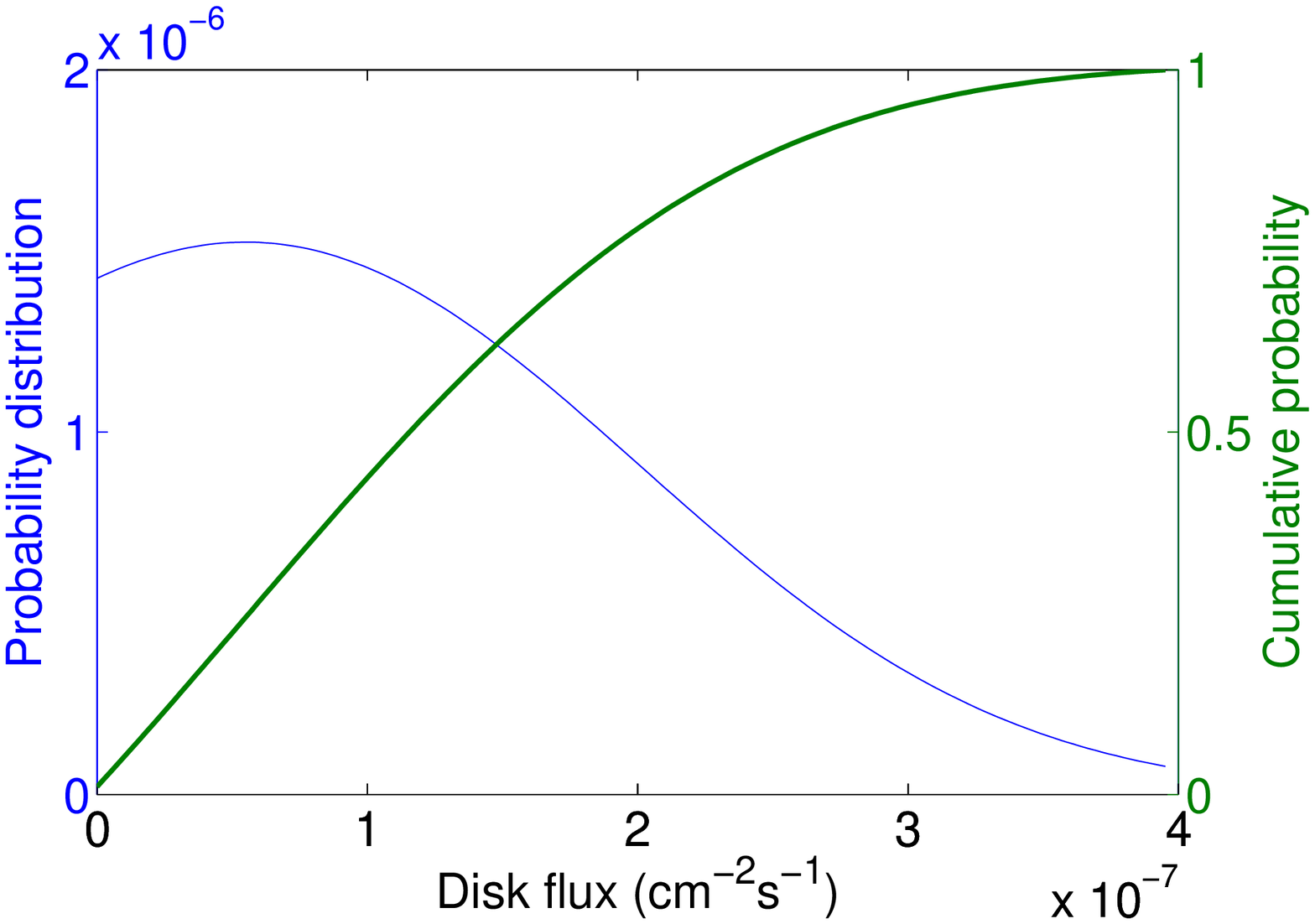}
	\caption{Marginal probability of the two components calculated from eq.\ref{eqA4} and cumulative probability as function of the extended (upper panel) and disk (lower panel) fluxes for different energy ranges, (left to right) above 100 MeV, above 300 MeV and for 100-300 MeV}
	 \label{fig13}
\end{figure*}

The log-likelihood ratio for the three energy ranges we analyzed are displayed in Fig.\ref{fig12} as a function of solar disk flux and extended flux.  Colors show different values of the ratio, obtained by allowing the background and the QSO 3C279 component to vary to maximize the likelihood for each value of disk flux and extended component, while contour lines define  1, 2 and 3$\sigma$  confidence intervals for the two separate parameters. 
Marginal probabilities of the two components calculated with the Bayesian method and cumulative probability as function of the disk and extended fluxes are shown in Fig. \ref{fig13}.
The sum of the two components, disk and extended, is given in Table \ref{table4}. 
\begin{table}
\caption{{\it Sum} of disk and extended components of solar emission for different energies. Values are from the Bayesian method.}             
\label{table4}      
\centering                          
\begin{tabular}{l l  }        
\hline\hline                 
Energy (MeV)& Total flux ($10^{-7}$ cm$^{-2}$s$^{-1}$)  \\    
\hline                        
   $> 100$ & (4.44$\pm$2.03) \\      

   $> 300$ &  (2.07$\pm$0.79)   \\

 $100-300$ &   (3.49$\pm$1.35) \\

\hline                                   
\end{tabular}
\end{table}
Counts of the source components centred on  the Sun resulting from the fitting technique above 300 MeV are shown in Fig.\ref{fig14}.
\begin{figure*}[!h]
 \centering
	\includegraphics[width=1.\textwidth, angle=0]{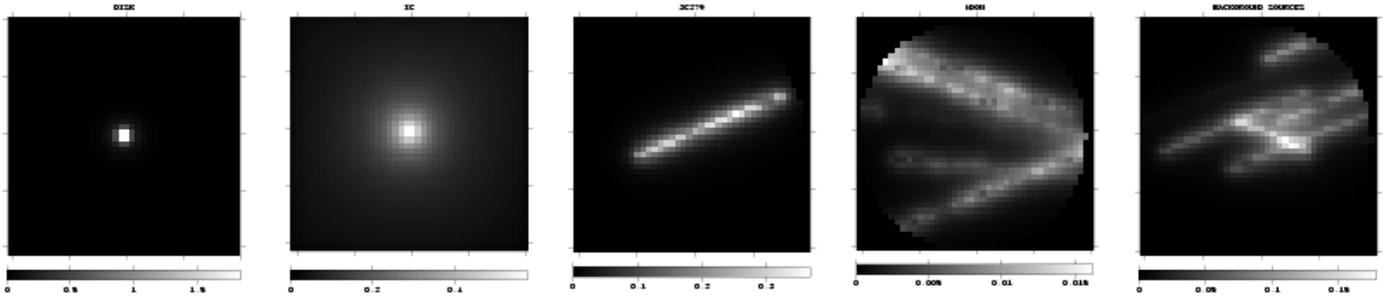}

	\caption{Counts of the source components resulting from the fitting technique above 300 MeV. Left to right: solar disk, inverse Compton, QSO 3C279, moon and background sources. Colorbars show counts per pixel.}
	 \label{fig14}
\end{figure*}
A summary of the main results is given in Table \ref{table5}. 
\begin{table}
\caption{Fluxes used to produce the plotted solar spectra. Fluxes are in  10$^{-7}$ cm$^{-2}$s$^{-1}$}
\label{table5}      
\centering          
\begin{tabular}{l l   l}     
\hline\hline       
Source  & 100-300 MeV &$>$300MeV  \\ 
\hline                    

\hline                    
Extended&2.1$\pm$1.3  & 1.7$\pm$0.9\\ 
Model extended & 1.3 & 0.9\\
Disk  & 1.4$\pm$0.9 &0.4$\pm$0.2 \\
Seckel's disk model &0-1.1 &0.1-0.5\\
\hline                  
\end{tabular}
\end{table}
The solar disk spectral index we found is 2.4$^{+0.9}_{-0.8}$. Errors are calculated from the uncertainties on the integrated fluxes. Figure (\ref{fig15}) shows the solar disk spectrum obtained imposing a constant spectral index of 2.4, its mean value, and following a simple power law up to 2 GeV.  
For the extended emission, we found a spectral index of 1.7$^{+0.8}_{-0.5}$.  The extended spectrum is shown in Fig. (\ref{fig16}). As for the disk spectrum, the spectral index has been fixed at the mean value. The spectrum is compared with the modelled spectrum for "naive" cases of 500 MV and 1000 MV solar modulation. 

\begin{figure}[!h]
 \centering
	\includegraphics[width=.5\textwidth, angle=0]{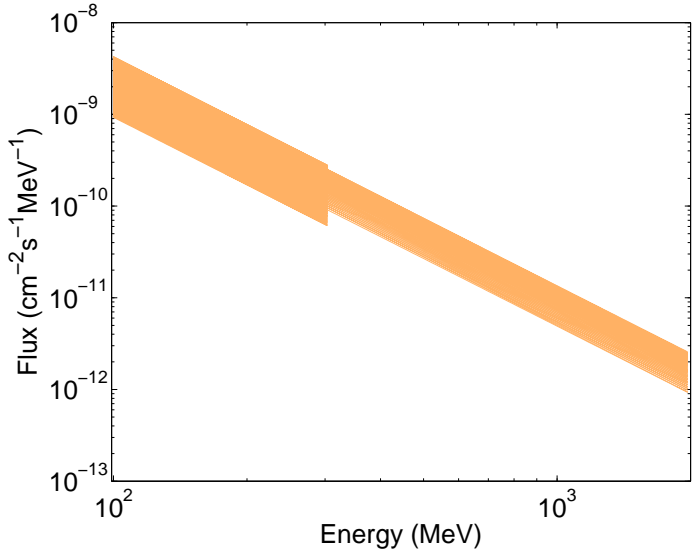}
	
	\caption{Solar disk spectrum. The orange regions defines the possible values obtained by varying the mean flux within 1$\sigma$ errors and for $\gamma$=2.4, the mean value of the spectral index.}
	 \label{fig15}
\end{figure}
\begin{figure}[!h]
 \centering
	\includegraphics[width=.5\textwidth, angle=0]{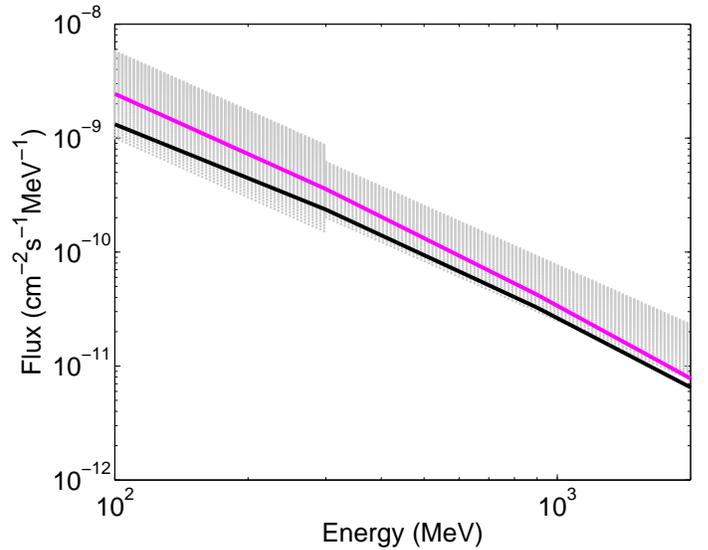}
	
	\caption{Solar extended spectrum. Gray regions define the possible values obtained by varying the mean flux within 1$\sigma$ errors and for $\gamma$=1.7, the mean value of the spectral index. Black line is the model for the naive case of 1000 MV modulation. For comparison, the pink line shows the model for the naive case of 500 MV modulation. }
	 \label{fig16}
\end{figure}

\section{Tests of analysis procedure}
In order to perform a test of the solar detection, we divided the total exposure which we used for the analysis described above, into two periods with about the same exposure time. Then, we analyzed the data with the same fitting technique explained above.
Both these time periods contain about 3.6$\times$10$^{8}$ cm$^{2}$s of exposure on the Sun.  Assuming equal periods and a constant flux from the Sun, the two periods should give compatible fit results.
As an example, the analysis was performed above 300 MeV, where the detection was most significant and for the naive case and solar modulation 1000 MV. 
The first period contains only 3C279, J1409-0745 and the moon, while in the second one there are only the other background sources and the moon; 3C279 is not present.
The total emission from the Sun was then detected at a level of 2.6$\sigma$ and 3.4$\sigma$ for first and second period respectively. The values obtained for the different emission components are consistent within 1$\sigma$. Moreover, they are also in agreement with those obtained for the total period. This confirms the validity of our method and the solar detection.

In order to exclude possible contamination of some faint solar flares detected by CGRO/COMPTEL during the same period, we performed the fitting method also with solar flares times excluded. Values will not be reported since, as expected, we did not find any change in the results.

\section{Discussion}

For the extended solar emission, Fig. (\ref{fig16}) shows that our model is fully consistent with the measured spectrum. 
The main uncertainty in the model is the electron spectrum (the solar radiation field and the physics of the inverse Compton are precisely known). The electron spectrum at the relevant energies has about a factor 2 experimental uncertainty. In fact the measured extended fluxes are higher by a factor $\sim$2 (although not very significant) which suggests our electron spectrum could be too low.

Regarding the solar disk flux, the flux in all cases is in agreement with the theoretical value for pion-decay obtained by \citet{seckel}.
The Galactic background obtained with the fitting technique is compatible with the expected Galactic emission (GALPROP) \citep{stronga}. The flux of 3C~279 of 7.2~$\times 10^{-7} cm^{-2}s^{-1}$, sum of the flux for 100-300 MeV and above 300 MeV, is in agreement with the EGRET catalogue. In the catalogue for period 11 used for this analysis the flux is ($7.94 \pm 0.75) \times 10^{-7} cm^{-2}s^{-1}$ above 100 MeV, while for period 12 (the other viewing period used in the analysis) there is hardly any contribution.

With respect to the analysis of EGRET data performed by \citet{thompson97}, our study has some improvements which explain why we succeeded in detection. Instead of excluding data near the sources 3C279 and the moon, they are included in our analysis. This leads to more exposure using contributions from all observing periods. The special dedicated analysis software for moving targets and the inclusion of  the extended emission from the Sun produced a more realistic prediction of the total data and hence a more sensitive analysis.

\section{Conclusions}
In this paper the theory of gamma-ray emission from IC scattering of solar radiation field by Galactic CR electrons has been given in detail. Analyzing the EGRET database, we find evidence for emission from the Sun and its vicinity. 
For all models the expected values are in good agreement with the data, with rather big errors bars. This also means that, because of the limited sensitivity of EGRET, it is not possible to prove which model better describes the data.
The spectrum of the moon has also been derived.

Since the inverse Compton emission from the heliosphere is extended, it contributes to the whole sky foreground and has to be taken into account for diffuse background studies. Moreover, since the emission depends on the electron spectrum and its modulation in the heliosphere, observations in different directions from the Sun can be used to determine the electron spectrum at different position, even very close to the Sun. To distinguish the modulation models (eg. with GLAST) an accuracy of $\sim$10$\%$ would be required. With the point source sensitivity of GLAST around 10$^{-7}$cm$^{-2}$s$^{-1}$ per day above 100 MeV, it will be possible to detect daily this emission, when the Sun is not close to the Galactic plane. In additional a precise model of the extended emission, such as that presented here, would be required in searches for exotic effects such as in \citet{fairbairn}

\begin{acknowledgements}

We thank Dirk Petry for providing counts maps, exposure and sources traces of the EGRET data and for useful discussions. We are also grateful to Berndt Klecker for useful discussions and drawing our attention to the Helios results. Thanks are due to the referee David Thompson for helpful suggestions which
significantly improved the presentation and analysis.
\end{acknowledgements}
\bibliographystyle{aa}

\Online
\begin{appendix} 
\section{Statistical method}

The likelihood statistic of binned data is the product of the probability of each pixel:
\begin{equation}
L=\prod_{ij} p_{ij}
\label{eqA1}
\end{equation}
where 
\begin{equation}
p_{ij}= \frac{\theta^{n_{ij}}_{ij}~e^{- \theta_{ij}}}{n_{ij}!}
\label{eqA2}
\end{equation}
is the Poisson probability of observing n$_{ij}$ counts in pixel (ij) where the number of counts predicted by the model is $\theta_{ij}$.
For each case analyzed, we obtained the logarithm of the likelihood ratio log~$(L/L_{0})$  where $L$ is the likelihood of the data with the all components included, and $L_{0}$ is the null hypothesis. We used both frequentist and Bayesian approach to analyze the results. 
Regarding the first method, -2~log~$(L/L_{m})$  is distributed as $\chi^{2}_{n}$, where $n$ is the number of parameters held fixed and $L_{m}$ is the global maximum \citep{strong85}. If 2 parameters are involved, we used the likelihood ratio test to determine the joint significance detection of the components, where -2~log~$(L/L_{0})$ is distributed as $\chi^{2}_{2}$. 
For the components separately it follows the $\chi^{2}_{1}$ distribution. 

In order to have a better treatment of our results, we performed also a Bayesian analysis treating likelihood as probability with uniform prior distribution \citep[see eg.][]{strong05}. This is independent of assumptions about distribution of the likelihood ratio. Hence, the mean value of one parameter is given by:
\begin{equation}
x_{io}=\int x_{i} P(x_{i})~ dx_{i}
\label{eqA3}
\end{equation}
where 
\begin{equation}
P(x_{i}) =\int_{i\neq j} P(\underline{x}) d^{n-1}x
\label{eqA4}
\end{equation}
is the marginal probability of x$_{i}$ with x$_{j}$ any value with i$\neq$j, n is the number of the parameters and
P($\underline{x}$) is the normalized joint probability distribution of $\underline{x}$, such that
\begin{equation}
\int P(\underline{x}) d^{n}x=1
\label{eqA5}
\end{equation}
The mean square error is then given by:
\begin{equation}
\Delta x_{i}^{2}=\int (x_{i}-x_{i0})^{2} P(x_{i}) dx_{i} 
\label{eqA6}
\end{equation}
When we are interested in the errors in the sum of the two components (i.e. disk and extended emission) we have to take into account their covariance, since they are correlated. Hence,
\begin{eqnarray}
(\Delta (x_{i}+x_{k}))^{2}=
(\Delta x_{i})^{2}+(\Delta x_{k})^{2}
\nonumber\\
+~2 \int (x_{i}-x_{i0})(x_{k}-x_{k0})~ P(\underline{x}) d^{n}x
\label{eqA7}
\end{eqnarray}

\end{appendix}

\end{document}